\documentclass[aps,pra,reprint, tightenlines,amsmath, amssymb,superscriptaddress]{revtex4-2}

\usepackage{graphicx}
\usepackage{epstopdf}
\usepackage{lipsum}
\usepackage[dvipsnames]{xcolor}
\usepackage[sort&compress]{natbib}
\usepackage{braket}

\pdfoutput=1
\usepackage[utf8]{inputenc}
\usepackage{bm}
\usepackage[english]{babel}
\usepackage[T1]{fontenc}
\usepackage{mathrsfs}
\usepackage[retainorgcmds]{IEEEtrantools}
\usepackage{amsmath}
\usepackage{amssymb}
\usepackage{color}
\usepackage{amsfonts}
\usepackage{times,txfonts}
\usepackage{nicefrac}
\usepackage{siunitx}
\usepackage[colorlinks=true,linkcolor=magenta,urlcolor=magenta,citecolor=magenta,pdfusetitle]{hyperref}

\usepackage{physics}
\usepackage{amssymb}

\usepackage{color}
\usepackage[dvipsnames]{xcolor}
\usepackage{comment}

\usepackage{subfiles}

\usepackage{amssymb,mathtools}
\usepackage{braket}

\usepackage{dsfont} 
\usepackage{color}
\def\bb{\begin{eqnarray}}
\def\ee{\end{eqnarray}}

\begin{document}
\title{Anomalous energy exchanges and Wigner function negativities in a single qubit gate}
\author{Maria Maffei}
\affiliation{Université Grenoble Alpes, CNRS, Grenoble INP, Institut Néel, 38000 Grenoble, France}

\author{Cyril Elouard}
\affiliation{Inria, ENS Lyon, LIP, F-69342, Lyon Cedex 07, France}

\author{Bruno O. Goes}
\affiliation{Université Grenoble Alpes, CNRS, Grenoble INP, Institut Néel, 38000 Grenoble, France}

\author{Benjamin Huard}
\affiliation{Ecole normale supérieure de Lyon, CNRS, Laboratoire de Physique, F-69342 Lyon, France}

\author{Andrew N. Jordan}
\affiliation{ Institute for Quantum Studies, Chapman University, 1 University Drive, Orange, CA 92866, USA}
\affiliation{Department of Physics and Astronomy, University of Rochester, Rochester, New York 14627, USA}

\author{Alexia Auffèves}
\affiliation{Université Grenoble Alpes, CNRS, Grenoble INP, Institut Néel, 38000 Grenoble, France}

\begin{abstract}
   
%Anomalous weak values and Wigner function's negativity are well known witnesses of quantum contextuality. We study a concrete setting where these effects manifest: a two-level emitter (qubit) resonantly coupled with a coherent waveguide field. This system implements a single qubit gate with a fidelity limited by the buildup of qubit-field correlations. We consider an experimental scheme typically implemented with superconducting circuits: the field is continuously monitored through heterodyne detection and then post-selected over the outcomes of a final qubit's measurement. The post-selected data can be interpreted as the field's weak values and can assume anomalous values. We model the joint system dynamics with a collision model, gaining access to the qubit-field entangled state at any time. We find an analytical expression of the quasi-probability distribution of the post-selected heterodyne signal, i.e. the conditional Husimi function. The latter grants access to all the field's weak values: we use it to obtain that of the field's energy change and display its anomalous behaviour. Finally, we derive the field's conditional Wigner function and show that anomalous weak values and Wigner function's negativities arise for the same values of the gate's angle.

Anomalous weak values and Wigner function’s negativity are well known witnesses of quantum contextuality. We show that these effects occur when analyzing the energetics of a single qubit gate generated by a resonant coherent field traveling in a waveguide. The buildup of correlations between the qubit and the field is responsible for bounds on the gate fidelity, but also for a nontrivial energy balance recently observed in a superconducting setup.
In the experimental scheme, the field is continuously monitored through heterodyne detection and then post-selected over the outcomes of a final qubit’s measurement. The post-selected data can be interpreted as field’s weak values and can show anomalous values in the variation of the field's energy. We model the joint system dynamics with a collision model, gaining access to the qubit-field entangled state at any time. We find an analytical expression of the quasi-probability distribution of the post-selected heterodyne signal, i.e. the conditional Husimi-Q function. The latter grants access to all the field’s weak values: we use it to obtain that of the field’s energy change and display its anomalous behaviour. Finally, we derive the field’s conditional Wigner function and show that anomalous weak values and Wigner function’s negativities arise for the same values of the gate’s angle.
\end{abstract}

\maketitle

\section{Introduction} 

Weak values have been originally defined as the average values for the results of weak measurements post-selected on particular outcomes of a final strong (projective) measurement~\cite{Aharonov1988}, where weak measurements are defined as measurements that minimally disturb the system~\cite{Wiseman2009book}. Later on, the concept of weak values has been generalized to any POVM (positive operator valued measurement) and any choice of the observable and the conditioning~\cite{Dressel2010}. 
When the outcome used for the post-selection is unlikely, weak values can exceed the range of eigenvalues of the corresponding operators~\cite{Aharonov1990}. In correspondence of such anomalous values, the Wigner function dictating the statistical distribution of the post-selected measurements takes negative values~\cite{Pusey2014}. Furthermore it can be proven that both anomalous weak values~\cite{Aharanov2002,Jordan2008,Pusey2014}, and Wigner function's negativity~\cite{Spekkens2008,Booth2021, Haferkamp2021} are witnesses of contextuality.

\begin{figure}[!htb]
\includegraphics[width = 0.5\textwidth]{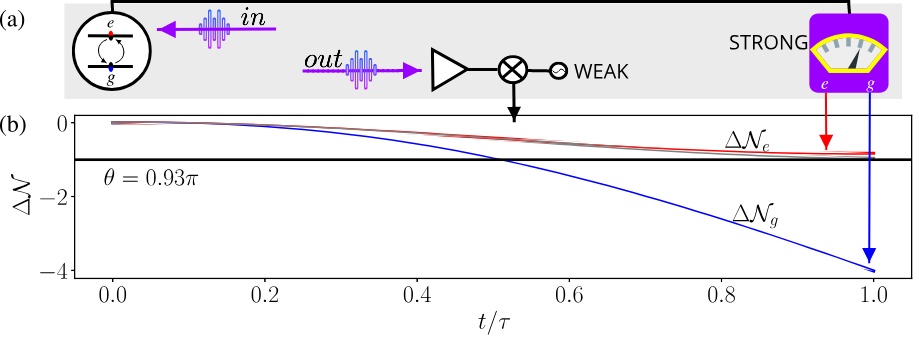}
\caption{Schematics of the detection of the field's weak values in the single-qubit gate. The gate is implemented by coherently driving a 1D atom with a pulse of area $\theta=\Omega\tau$; the output field is continuously detected in the time interval $[0,\tau]$ with a heterodyne measurement (weak measurement); at time $\tau$ a projective (strong) measurement is performed on the qubit and the heterodyne data acquired are post-selected according on the outcome. The plot shows the values of the change of the number of field's excitations as a function of time for $\theta=0.93 \pi$ and $\gamma\tau = 3/40$. The gray line represents the unconditional change of the number of excitations.}
\label{fig:fig1}
\end{figure}
Here we show that those effects occur in a paradigmatic setting of waveguide quantum electrodynamics where energy exchanges feature anomalous weak values. The setting is the so called one-dimensional (1D) atom, a two-level emitter (qubit) interacting with an electromagnetic field propagating in 1D. When the field is prepared in a coherent state resonant with the qubit's transition this system implements a single qubit gate with a fidelity limited by the buildup of qubit-field correlations~\cite{Bertet2001}. Weak values arise in the 1D atom when the electromagnetic field is monitored via heterodyne detection and the data are post-selected over the outcomes of a qubit's projective measurement. This may be understood by regarding the propagating field as a weak measurement apparatus for the qubit~\cite{Wiseman2002,Dressel2014}, see Fig.~\ref{fig:fig1}(a). Superconducting circuits represent the ideal setup to implement such a detection scheme as they grant independent access to the states of qubit and field. A recent experiment on a superconducting single qubit gate~\cite{Stevens2022} showed anomalous weak values of the field's energy change, i.e. values exceeding by far the single quantum of energy that qubit and field can physically exchange, see Fig.~\ref{fig:fig1}(b).

We study the 1D atom using a collision model~\cite{Ciccarello2020} where individual temporal modes of the electromagnetic field locally interact with the qubit in a sequential fashion~\cite{Ciccarello2017,Gross2018}. When the field is prepared in a resonant coherent state, this method provides the exact analytical expression of the qubit-field entangled state at any time~\cite{Maffei2022}. From the collision model of the driven 1D atom, we derive the analytical expression of the field's Husimi-Q function conditioned on the outcomes of the final qubit's measurement. This function gives access to all the moments of the post-selected heterodyne distribution, namely the field's weak values. Then we use the conditional Husimi-Q function to derive the weak value of the field's energy change and we show that, in the typical working regime of single qubit gates, it takes anomalous values as observed in~\cite{Stevens2022}. Finally we explore the relation between anomalous weak values and negativity of the corresponding Wigner function. Exploiting the analytical expression of the field's state obtained with the collision model, we compute the field's conditional Wigner function and we show that anomalous weak values and Wigner function's negativities arise for the same values of the gate's angle.

\textcolor{black}{
 The paper is organized as follows: In Sec.~\ref{Sec2} we describe the coherently driven 1D atom and present the collision model of its dynamics. In Sec.~\ref{sec:FieldWeakValues} we present our main results: the analytical derivations of the field's conditional Husimi-Q function and the weak value of the field's energy change. In Sec.~\ref{ConditionalWFunction} we derive the conditional Wigner function. Finally, in Sec.~\ref{Conclusions} we draw the conclusions of our work.  
 }
\section{System and model}\label{Sec2}
\subsection{The coherently driven 1D atom}
The 1D atom comprises a \textcolor{black}{qubit} coupled to a single-mode, semi-infinite waveguide. The waveguide field constitutes a reservoir of electromagnetic modes of frequencies $\omega_k$ and linear momentum $k=\omega_k v^{-1}$ with $v$ being the field's group velocity taken as positive. These modes are destroyed (created) by the operators $a_k(a^{\dagger}_k)$. The dynamics of the joint system is ruled by the Hamiltonian:
\begin{align}
H & = \left[\hbar\omega_{0}\sigma^{\dagger}\sigma+\hbar\sum_{k=0}^{\infty}\omega_{k}a_{k}^{\dagger}a_{k}\right]+i\hbar g\sum_{k=0}^{\infty}\left(\sigma^{\dagger}a_{k}-a_{k}^{\dagger}\sigma\right), \label{eq:normal-Hamiltonian} 
\end{align}
where $\sigma \equiv \ket{g}\bra{e}$, with $\ket{g(e)}$ being the ground (excited) state of the emitter. Writing the above Hamiltonian we implicitly assumed that the light-matter interaction is weak enough that only frequency modes close to $\omega_0$ play a role (quasi-monochromatic approximation)~\cite{Gross2018}. In this regime, the rotating wave approximation is allowed~\cite{Loudon2000}, and the coupling $g$ can be considered uniform in frequency~\cite{Gardiner1985}.

The field's lowering operator at the position $x$ in the interaction picture~\cite{Ciccarello2017,Gross2018,Maffei2022} is given by
\begin{equation} 
\label{eq:a}
b(x,t)=\sqrt{\frac{1}{\varrho}}\sum_{k} e^{-i \omega_k( t-x/v)}a_k=b(0,t-x/v),
\end{equation}
where $\varrho$ is the modes' density verifying the relation $\sum_{k}e^{-i\omega_{k}(t-t')}/\varrho=\delta(t-t')$. The operators $b(0,t)$ satisfy the bosonic commutation relation, i.e. $[b(0,t),b^{\dagger}(0,t')]=\delta(t-t')$~\cite{Gross2018}. The qubit is located at the position $x=0$ of the waveguide, such that the interaction picture Hamiltonian reads:
\begin{equation}
H_{I}(t)=i\hbar\sqrt{\gamma}\left(\sigma^{\dagger}(t)b(0,t)-b^{\dagger}(0,t)\sigma(t)\right),\label{eq:Int-pic-Hamiltonian}
\end{equation}
where we defined the interaction picture lowering operator $\sigma(t)=e^{-i\omega_{0}t}\sigma$, and the emitter's decay rate $\gamma=g^{2}\varrho$. In the regions where $x<0$ or $x>0$, the field travels without deformation, it is then natural to define field's input and output operators respectively as $b_{\text{in}}(t)\equiv\lim_{\epsilon\rightarrow0^{-}}b(\epsilon,t)$, and $b_{\text{out}}(t)\equiv\lim_{\epsilon\rightarrow0^{+}}b(\epsilon,t)$. These operators satisfy the mean input-output relation $\langle b_{\text{out}}(t)\rangle=\langle b_{\text{in}}(t)\rangle-\sqrt{\gamma} \langle \sigma(t)\rangle$, in agreement with the textbook input-output relation written in the Heisenberg representation~\cite{Gardiner1985}. 

The gate's input field is a square coherent pulse of amplitude $\langle b_{in}(t)\rangle=\alpha_t=\alpha e^{-i\omega_0 t}/\sqrt{\varrho}$, with $\alpha$ real. Hence, the field's state at the initial time ($t=0^{-}$) reads $\ket{\alpha}\equiv \mathcal{D}(\alpha)\ket{0}$, where $\mathcal{D}(\alpha)=e^{ \left(\alpha a^{\dagger}_0-\alpha^{*}a_0\right)}$ is the displacement operator of the mode with frequency $\omega_0$ which can be equivalently written as $\mathcal{D}(\alpha)=e^{\int dt \left(\alpha_t b^{\dagger}(t)-\alpha^{*}_tb(t)\right)}$ using the transformation~\eqref{eq:a}. A unitary driving on the qubit, $H_D(t)$, arises naturally when displacing the interaction Hamiltonian in Eq.~\eqref{eq:Int-pic-Hamiltonian}, $\mathcal{D}(-\alpha)H_{I}(t)\mathcal{D}(\alpha)=H_{I}(t)-\Omega\sigma_y/2=H_{I}(t)+H_{D}(t)$, with $\Omega/2=g\alpha$ and $\sigma_y=i\sigma^{\dagger}-i\sigma$. In the classical limit of the field~\cite{Bertet2001}, the qubit reduced dynamics is solely dictated by $H_{D}(t)$ and hence it reduces to a pure rotation around the $y$ axis of an angle $\theta=\Omega \tau$, where $\tau$ is the duration of the qubit-field interaction. In this limit, the light-matter interaction (int) is equivalent to the map:
\begin{align}\label{eq:map}\nonumber
    \ket{g}\otimes\ket{\alpha}\xrightarrow{\text{int}(\tau)}\left[\cos{(\theta/2)}\ket{g}+\sin{(\theta/2)}\ket{e}\right]\otimes\ket{\alpha}.
\end{align}

Beyond the classical limit, the interaction entangles qubit and field resulting in a loss of purity of the reduced qubit's state and a degradation of the coherence of the input field. The joint qubit-field state at time $\tau$ can be written as the pure state:
\begin{align}\nonumber
    \ket{\Psi(\tau)}=\sqrt{P_{g}(\tau)}\ket{g,\psi_g(\tau)}+\sqrt{P_{e}(\tau)}\ket{e,\psi_e (\tau)}
\end{align}
with
\begin{equation}\label{eq:Fwavefunction}
\ket{\psi_{\epsilon}(\tau)} =\mathcal{D}(\alpha)\left(\frac{f_{\epsilon}^{(0)}(\tau)-\int_{0}^{\tau} dt f^{(1)}_{\epsilon}(\tau,t)e^{-i\omega_0 t} b^{\dagger}(t)+\dots}{\sqrt{P_{\epsilon}(\tau)}}\right)\ket{0},
\end{equation}
where $\epsilon=g,e$, and we adopted the short notation $b(t)\equiv b(0,t)$. The state in the parenthesis is the field state in the displaced reference frame, where the interaction Hamiltonian is $\mathcal{D}(-\alpha) H_{I}(t)\mathcal{D}(\alpha)$, and the input field is the vacuum. In such a frame, the only mechanism responsible for the photons' creation is the spontaneous emission, hence the field coincides with the emitter's fluorescence having amplitude $\langle b_{out}(t)\rangle-\langle b_{in}(t)\rangle=-\sqrt{\gamma}\langle \sigma(t)\rangle$. The ellipsis represent the components with $j>1$ spontaneously emitted photons, having the form $(-)^{j}\int d\textbf{t}_j f_{\epsilon}^{(j)}(\tau,\textbf{t}_j)e^{-i\omega_0 t_1}b^{\dagger}(t_1)..e^{-i\omega_0 t_j}b^{\dagger}(t_j)\ket{0}$, with $\textbf{t}_j=\lbrace t_1,t_2,...t_j\rbrace$. The functions $f_{\epsilon}^{(j)}(\tau,\textbf{t}_j)$ are real and their explicit expression has been derived in~\cite{Maffei2022} and is reported in the App.~\ref{App:FieldsWavefunction}. Since their amplitude is proportional to $\gamma^{j/2}$, the components with $j>2$ can be neglected in the regime usually considered in single-qubit gates where $\Omega\gg \gamma$. We can define the probability that the qubit spontaneously emits $j$ photons during the evolution from $\ket{g}$ to $\ket{\epsilon}$ as $p^{(j)}_{\epsilon}(\tau)\equiv\int d\textbf{t}_j |f_{\epsilon}^{(j)}(\tau,\textbf{t}_j)|^2/P_{\epsilon}(\tau)$.

\subsection{Collision model of the coherently driven 1D atom}

We now define adimensional discrete-temporal modes of the electromagnetic field~\cite{Gross2018,Ciccarello2017,Maffei2022}, i.e. $b_n\equiv\sqrt{\Delta t}b(t_n)$, where $\Delta t$ is an infinitesimal time-increment, $n$ is an integer number, and $[b_n,b^{\dagger}_{n'}]=\delta_{n,n'}$ with $\delta_{n,n'}$ being the Kronecker delta~\cite{Ciccarello2017,Gross2018}.

The initial state of the field can be written in terms of the discrete-temporal modes as: $\ket{\alpha}=\bigotimes_{n} \ket{\alpha_n}$, where $\ket{\alpha_n}=\mathcal{D}^{(n)}(\alpha_n)\ket{0_n}$, with $\mathcal{D}^{(n)}(\alpha_n)=e^{(\alpha_n b^{\dagger}_n-\alpha_n^* b^{\dagger}_n)}$, and $\alpha_n=\sqrt{\Delta t/\varrho}\alpha e^{-i\omega_0 t_n}$.

We can write the infinitesimal unitary evolution operator $U(t_{n+1}-t_{n})$ as:
\begin{align}
U^{(n)}&\equiv U\left(t_{n+1}-t_{n}\right)\approx\exp\left\lbrace -\frac{i}{\hbar}\Delta t H_{I}\left(t_{n}\right)\right\rbrace,\mbox{with }\\ \nonumber
H_{I}(t_n) &\equiv i\hbar\sqrt{\frac{\gamma}{\Delta t}}\left(\sigma^{\dagger}(t_n)b_n-b^{\dagger}_n\sigma(t_n)\right).
\end{align}
The joint system evolution can be decomposed in a sequence of collisions:
\begin{align}\label{eq:rho}
    \rho(t_{n+1})=U^{(n)}\rho(t_{n})U^{(n)\dagger}.
\end{align}
The qubit's reduced state at time $t_{n+1}$ can be found by tracing Eq.~\eqref{eq:rho} over the state of the $n$-th temporal mode prepared in the state $\ket{\alpha_n}$:
\begin{align}\label{eq:qevol}
    \rho_{q}(t_{n+1})=\text{Tr}_{n}\lbrace U^{(n)}\rho_{q}(t_{n}) \ket{\alpha_n}\bra{\alpha_n}U^{(n)\dagger}\rbrace.
\end{align}
As expected, expanding $U^{(n)}$ at the second order in $\Delta t$, equation~\eqref{eq:qevol} becomes a discrete-time Lindblad master equation~\cite{Ciccarello2017, Gross2018}:
\begin{align}
    \frac{\rho_\text{q}(t_{n}) -\rho_\text{q}(t_{n-1})}{\Delta t} &= \frac{\Omega}{2}  [\sigma-\sigma^{\dagger},\rho_\text{q}(t_{n-1})]\\ \nonumber
    &+ \gamma\Big( \sigma \rho_\text{q}(t_{n-1})\sigma^{\dagger} - \tfrac{1}{2}\Big\lbrace\sigma^{\dagger}\sigma,\rho_\text{q}(t_{n-1})\Big\rbrace \Big).
\end{align}
Notice that at time $t_{n}$, the $(n-1)$-th discrete-temporal mode of the field has just interacted with the qubit, while the $n$-th is going to interact next. Then, input and output operators correspond to the limits: 
\begin{align}\label{eq:inout}
 b_{in}(t_n)&=\lim_{\Delta t\rightarrow 0}\frac{b_n}{\sqrt{\Delta t}},\\ \nonumber b_{out}(t_n)&=\lim_{\Delta t\rightarrow 0}\frac{b_{n-1}}{\sqrt{\Delta t}}.   \end{align}

\section{Field weak values}\label{sec:FieldWeakValues}

The weak value of an operator at time $t$ can be written as~\cite{Dressel2010,Dressel2012,Dressel2014}: 
\begin{align}\label{eq:weak_O}
    \langle O(t)\rangle_{f,i}=\frac{\text{Tr}\lbrace \Pi_{f} U(\tau-t)O(t)\rho_{i}(t)U^{\dagger}(\tau-t)\rbrace}{P_{f}(\tau)},
\end{align}
 where $\Pi_{f}=\ket{f}\bra{f}$ is the projector on the final measurement's outcome, $U(\tau-t)=\text{exp}\lbrace (-i/\hbar) \int_{t}^{\tau} dt' H_{I}(t')\rbrace$, and $\rho_{i}(t)=U(t)\ket{i}\bra{i}U^{\dagger}(t)$ is the system's state at time $t$ starting from the pure state $\ket{i}$. From now on we will consider that the joint qubit-field system starts its evolution from the state $\ket{g,\alpha}$, and that the qubit is measured in its energy basis at the final time $\tau$. Then, in the rest of the paper, we will simplify the weak values' notation by omitting the subscript that refers to the initial state, and by using $\epsilon=e,g$ to denote the possible outcomes of the qubit's final measurement.

We are particularly interested in the weak values of the output field's quadratures, $\text{Re}\lbrace \langle b_{out}(t)\rangle_{\epsilon}\rbrace$ and $\text{Im}\lbrace \langle b_{out}(t)\rangle_{\epsilon}\rbrace$, and intensity, $\langle b^{\dagger}_{out}(t)b_{out}(t)\rangle_{\epsilon}$. Indeed, these quantities can be measured by implementing the driven 1D atom with a superconducting circuit, performing heterodyne detection on the output field, and postselecting it on the outcomes of the final qubit's measurement as in Refs.~\cite{Campagne2014,Stevens2022}.

Furthermore, from the weak value of the output intensity, we can derive the weak value of the change of the field's number of excitations:
\begin{align}\label{eq:weak_n_def}
\Delta \mathcal{N}_{\epsilon}&\equiv\int_{0}^{\tau} dt\langle b_{out}^{\dagger}(t)b_{out}(t)\rangle_{\epsilon}-|\alpha_t|^2\tau.
\end{align}
In the typical regime used to perform single qubit gates, i.e. $\Omega\gg\gamma$, this quantity may take anomalous values, namely values exceeding by far the single quantum of excitation that the field can physically exchange with the qubit, i.e. $|\Delta \mathcal{N}_{\epsilon}|>1$, see Fig.~\ref{fig:fig1}(b), and Refs.~\cite{Stevens2021,Rogers2022}.

Using the collision model picture introduced in the previous section, we can derive explicit expressions of the weak values of any field's operator of the kind $O(t)=\left(b^{\dagger}_{out}(t)\right)^m \left(b_{out}(t)\right)^{l}$, hence including output quadratures and intensity. The first step is to write these weak values in terms of the discrete-time output operator (Eq.~\eqref{eq:inout}):
\begin{widetext}
\begin{align}
    \langle\label{eq:bbnweak} \left(b^{\dagger}_{out}(t_n)\right)^m \left(b_{out}(t_n)\right)^{l}\rangle_{\epsilon}&=\lim_{\Delta t\rightarrow 0} \frac{\text{Tr}\lbrace \Pi_{\epsilon} U(\tau-t_{n})\left(b^{\dagger}_{n-1}\right)^m b_{n-1}^{l}\rho(t_{n})U^{\dagger}(\tau-t_{n})\rbrace}{\sqrt{\Delta t}}\\\nonumber
    &=\lim_{\Delta t\rightarrow 0} \frac{\bra{\psi_{\epsilon}(\tau)}\left(b^{\dagger}_{n-1}\right)^m b_{n-1}^{l}\ket{\psi_{\epsilon}(\tau)}}{\sqrt{\Delta t}},
\end{align}
\end{widetext}
where we used the fact that $b_{n-1}$ commutes with $U(\tau-t_n)$.
Equation~\eqref{eq:bbnweak} shows that the weak value of an arbitrary observable of the output field is simply its average value on the state $\ket{\psi_{\epsilon}(t)}$ given in Eq.~\eqref{eq:Fwavefunction}.

Hence, to evaluate Eq.~\eqref{eq:bbnweak} we can use the conditional Husimi-Q function of the $n$-th temporal mode:
\begin{align}\label{eq:QF}
    &\mathcal{Q}^{(n)}_{\epsilon}(s)\equiv \frac{1}{\pi}\text{Tr}\lbrace \Pi^{(n)}_{s}\ket{\psi_{\epsilon}(\tau)}\bra{\psi_{\epsilon}(\tau)}\rbrace,
\end{align}
 where $\Pi^{(n)}_{s}=\ket{s_{n}}\bra{s_{n}}$ with $\ket{s_{n}}=\mathcal{D}^{(n)}(s)\ket{0_{n}}$ and $\mathcal{D}^{(n)}(s)=e^{(s b^{\dagger}_n-s^* b^{\dagger}_n)}$ being the displacement operator of the $n$-th temporal mode. From the Husimi function it is possible to obtain any moment of the output field's distribution just performing an integral in the complex plane~\cite{Walls2008representations}. So, for instance, the weak value of the output intensity at time $t_n$ can be obtained from $\mathcal{Q}^{(n)}_{\epsilon}(s)$ by doing the integral:
 \begin{align}\label{eq:numbt}
    \langle b^{\dagger}_{out}(t_n) b_{out}(t_n) \rangle_{\epsilon}&=\lim_{\Delta t\rightarrow 0}\frac{1}{\Delta t}\int d^{2}s (|s|^2-1)\mathcal{Q}^{(n)}_{\epsilon}(s).
\end{align}

%\begin{align}\label{eq:weak_O}
    %\langle\label{eq:bnweak} b_{out}(t_{n})\rangle_{\epsilon}&=\lim_{\Delta t\rightarrow 0} \frac{\text{Tr}\lbrace \Pi_{\epsilon} U(\tau-t_{n})b_{n-1}\rho(t_{n})U^{\dagger}(\tau-t_{n})\rbrace}{\sqrt{\Delta t}}\\\nonumber&=\lim_{\Delta t\rightarrow 0} \frac{\bra{\psi_{\epsilon}(\tau)}b_{n-1}\ket{\psi_{\epsilon}(\tau)}}{\sqrt{\Delta t}},
%\end{align}
 
%\begin{align}
%\langle b_{out}(t_n) \rangle_{\epsilon}&=\lim_{\Delta t\rightarrow 0}\frac{1}{\sqrt{\Delta t}}\int d^{2}s  s \mathcal{Q}^{(n)}_{\epsilon}(s),
%\end{align}

The explicit expression of $\mathcal{Q}^{(n)}_{\epsilon}(s)$ in terms of qubit's operators can be computed using the collision model. This is the main result of this section. First, we rewrite Eq.~\eqref{eq:QF} as:
\begin{align}\label{eq:QFE}
    \mathcal{Q}^{(n)}_{\epsilon}(s)=\frac{1}{\pi P_{\epsilon}(\tau)}\text{Tr}\lbrace E_{\epsilon}(\tau,t_{n+1})\Pi^{(n)}_{s}\rho(t_{n+1})\Pi^{(n)}_{s}\rbrace,
\end{align}
where $E_{\epsilon}(\tau,t)\equiv U^{\dagger}(\tau-t)\ket{\epsilon}\bra{\epsilon}U(\tau-t)$ is the so-called effect matrix~\cite{Gemmelmark2013}. Now, plugging Eq.~\eqref{eq:qevol} in Eq.~\eqref{eq:QFE}, and expanding $U^{(n)}$ at the second order in $\Delta t$, we find:
\begin{widetext}
\begin{align}
{\cal Q}_{\epsilon}^{(n)}(s) & =\frac{1}{\pi P_{\epsilon}(\tau)}\text{Tr}\left\{ E_{\epsilon}(\tau,t_{n+1})\langle s_{n}\vert U^{(n)}\vert\alpha_{n}\rangle\rho_{q}(t_{n})\langle\alpha_{n}\vert\left(U^{(n)}\right)^{\dagger}\vert s_{n}\rangle\right\} \\
 & =\text{Tr}\left\{ E_{\epsilon}(\tau,t_{n+1})\left[\rho_{q}(t_{n+1})+\gamma\Delta t\left(\vert\alpha_{n}-s\vert^{2}-1\right)\sigma\rho_{q}(t_{n})\sigma^{\dagger}+\sqrt{\gamma\Delta t}\left((\alpha_{n}-s)^{*}\sigma(t_{n})\rho_{q}(t_{n})+(s-\alpha_{n})\rho_{q}(t_{n})\sigma^{\dagger}(t_{n})\right)\right]\right\} \frac{\exp\left\{ -\vert s-\alpha_{n}\vert^{2}\right\} }{\pi P_{\epsilon}(\tau)}\nonumber
\end{align}
\end{widetext}
%\begin{align}\nonumber
%    \mathcal{Q}^{(n)}_{\epsilon}(s)&=\frac{\text{Tr}\left\lbrace E_{\epsilon}(\tau,t_{n+1})\bra{s_n}U^{(n)}%%\ket{\alpha_n}\rho_{q}(t_{n})\bra{\alpha_n}U^{(n)\dagger}\ket{s_n}\right\rbrace}{\pi P_{\epsilon}(\tau)}\\ %\nonumber
%    &=\text{Tr} \left\lbrace E_{\epsilon}(\tau,t_{n+1})\left[\rho_{q}(t_{n+1})+\gamma\Delta t (|\alpha_n-s|^2-1)\sigma\rho_{q}(t_n)\sigma^{\dagger}\right\rbrace.\\ \nonumber &\left.\left.\left. +\sqrt{\gamma\Delta t}%\left((\alpha_n-s)^*\sigma(t_n) \rho_{q}(t_n)+(s-\alpha_n) \rho_{q}(t_n)\sigma^{\dagger}(t_n)\right)\right]%\right\rbrace \frac{e^{-|s-\alpha_n|^2}}{\pi P_{\epsilon}(\tau)}.
%\end{align}
Noticing that $\text{Tr}\lbrace E_{\epsilon}(\tau,t_{n+1})\sigma(t_n) \rho_{q}(t_n)\rbrace/P_{\epsilon}(\tau)=\langle \sigma(t_n)\rangle_{\epsilon}$~\cite{Gemmelmark2013}, and that $\text{Tr}\lbrace E_{\epsilon}(\tau,t_{n+1}) \rho_{q}(t_{n+1})\rbrace=P_{\epsilon}(\tau)$, we find: 
\begin{align}\label{eq:QFEfinal}
     \mathcal{Q}^{(n)}_{\epsilon}(s)&=\frac{e^{-|s-\alpha_n|^2}}{\pi}\left[1+ \left(|\alpha_n-s|^2-1\right)\gamma \Delta t \mathcal{J}_{\epsilon}(t_n)\right.\\ \nonumber
     &\left.+2 \sqrt{\gamma \Delta t}\text{Re}\lbrace(s-\alpha_n) \langle \sigma(t_n)\rangle_{\epsilon}\rbrace \right],
\end{align}
where we defined $\gamma \mathcal{J}_{\epsilon}(t_n)=\text{Tr}\lbrace \Pi_{\epsilon}U(\tau-t_n) \sigma \rho_\text{q}(t_{n})\sigma^{\dagger}U^{\dagger}(\tau-t_n)\rbrace/P_\epsilon(\tau)$. An alternative, although equivalent, derivation of the conditional Husimi-Q function from the wavefunction in Eq.~\eqref{eq:Fwavefunction} is reported in the App.~\ref{App:HusimiDerivation}.

Uning Eq.~\eqref{eq:QFEfinal} to perform the integral in Eq.~\eqref{eq:numbt}, and then plugging the result in Eq.~\eqref{eq:weak_n_def}, we find:
\begin{align}\label{eq:weak_n_sigma}
\Delta \mathcal{N}_{\epsilon}=\int_{0}^{\tau}dt\left(\gamma \mathcal{J}_{\epsilon}(t)-\Omega\text{Re}\lbrace \langle \sigma(t)\rangle_{\epsilon}\rbrace \right).
\end{align}
Notice that $\int_{0}^{\tau} dt \gamma \mathcal{J}_{\epsilon}(t)=\sum_{j\geq1} p^{(j)}_{\epsilon}(\tau)/P_{\epsilon}(\tau)$ is the total probability that the qubit undergoes spontaneous emission along its evolution from $\ket{g}$ to $\ket{\epsilon}$. While the last term, $-\int_{0}^{\tau}dt \Omega\text{Re}\lbrace \langle \sigma(t)\rangle_{\epsilon}\rbrace=2\int_{0}^{\tau}dt\text{Re}\lbrace \langle b_{in}(t)\rangle^{*}\left(\langle b_{out}(t)\rangle_{\epsilon}-\langle b_{in}(t)\rangle\right)\rbrace,$ features an interference between the input field and the emitter's fluorescence post-selected over the outcome $\epsilon$; this term is the only part of $\Delta\mathcal{N}_{\epsilon}$ whose modulus can exceed $1$ leading to anomalous values.

%Figure~\ref{fig:fig1} illustrates schematically the measurement scheme of the field's weak values $\Delta \mathcal{N}_{\epsilon}$. The average number of photons in the output field is continuously detected via an heterodyne measurement from 0 to $\tau$, then the data are post-selected according on the outcome of the final qubit's projective measurement made at $t=\tau$. The plot shows the values of $\Delta \mathcal{N}(t)$ (unconditional change in the field's number of excitation) $\Delta \mathcal{N}_{g}(t)$, and $\Delta \mathcal{N}_{e}(t)$ expected for a gate with $\theta=\Omega \tau=0.85\pi$, and $\gamma^{-1}=8\tau$. In this regime, the average qubit's state at $t=\tau$, is close to the excited state, and, accordingly, the corresponding gate's field looses almost one excitation in average, $\Delta \mathcal{N}(\tau)\approx -1$. The plot shows that $\Delta \mathcal{N}_{g}(t)<\Delta \mathcal{N}(t)$, while $\Delta \mathcal{N}_{e}(t)>\Delta \mathcal{N}(t)$. This behavior can be understood as an informational back-action of the qubit's projective measurement on the field's state~\cite{Stevens2021}: measuring $\ket{g}$ projects the qubit backwards with respect to its average evolution, and the corresponding gate's field onto a state with less energy than its expected value; vice versa, measuring $\ket{e}$ projects the qubit ahead, and the corresponding gate's field onto a more energetic state.

\section{Conditional Wigner function}\label{ConditionalWFunction}

\begin{figure}[!htb]
\includegraphics[width = 0.5\textwidth]{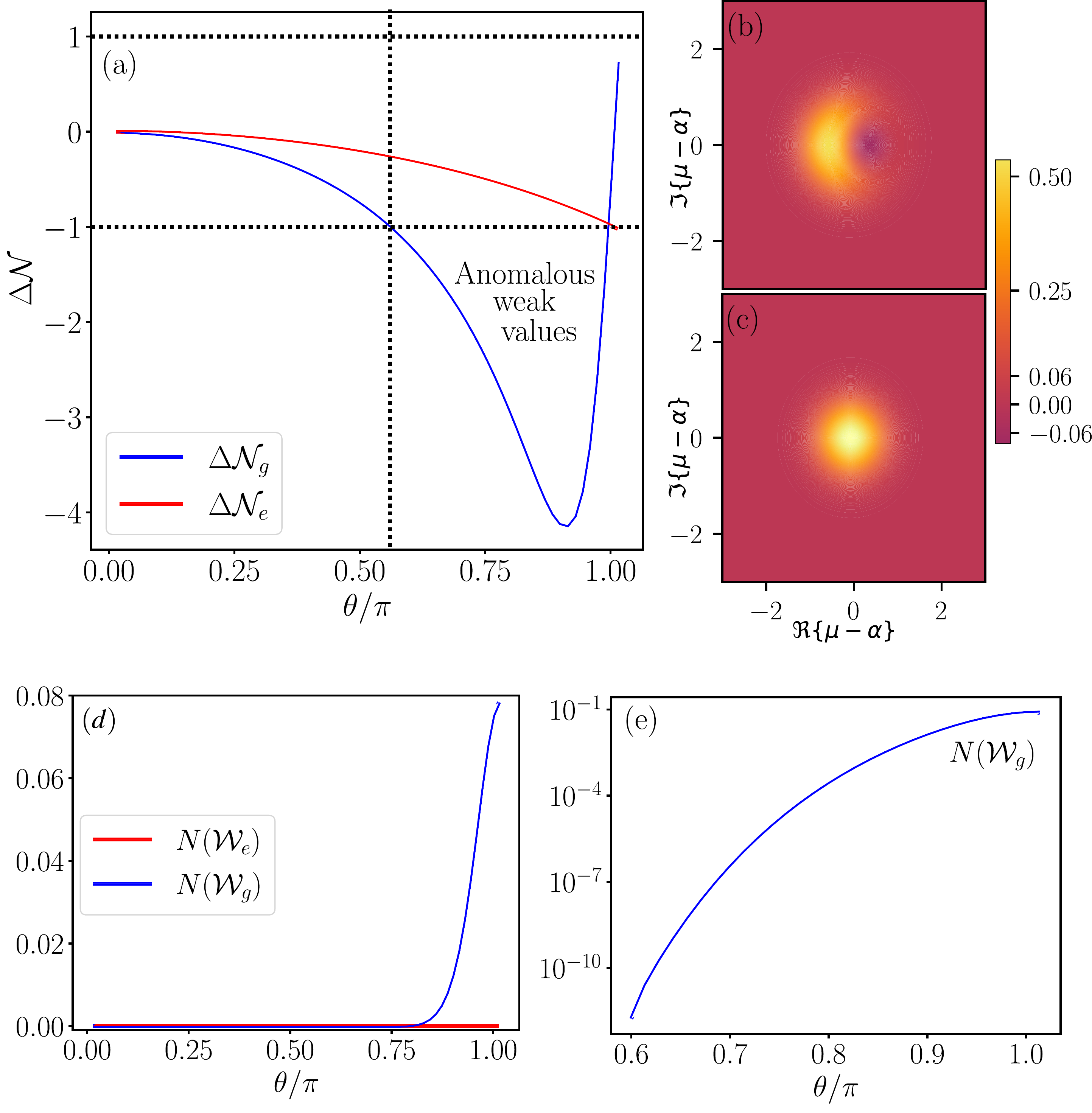}
\caption{Anatomy of the single-qubit gate with  $\gamma\tau = 3/40$, varying the gate's angle $\theta$. (a) Weak values of the change of number of excitations in the field when the qubit at time $\tau$ is found in $\ket{\epsilon}$ with $\epsilon=e,g$, $\Delta \mathcal{N}_{\epsilon}$. (b-c) Color plots of $\mathcal{W}_{g}$ and $\mathcal{W}_{e}$ for $\theta/\pi = 0.93$. (d) Wigner function's negativity, $N(\mathcal{W}_{\epsilon})=\int d^2\mu |\mathcal{W}_{\epsilon}(\mu)|-1$, as a function of $\theta$: $\mathcal{W}_{e}$ is always positive, while $\mathcal{W}_{g}$ can take negative values in the region of anomalous weak values of $\Delta \mathcal{N}_{g}$. (e) Plot of $\mathcal{W}_{g}$ (log. scale) in the region of anomalous values of $\Delta \mathcal{N}_{g}$, i.e. $\theta\in[0.6\pi,\pi]$.}
\label{fig:fig2}
\end{figure}
In the previous section, we derived the conditional Husimi-Q function of the field's temporal modes to predict the results of a continuous heterodyne detection with post-selection. Here instead we consider the mode of frequency $\omega_0$, being the qubit's frequency and the center of the field's spectrum. In the quasi-monochromatic, resonant regime usually employed in single qubit gates, $\omega_0$ is much more populated than the other frequencies, so $\mathcal{N}(\omega_0)\equiv \langle a^{\dagger}_0 a_0\rangle \approx \int_{0}^{\tau} dt \langle b^{\dagger}(t) b(t)\rangle$, and consequently $ \Delta\mathcal{N}_{\epsilon}(\omega_0)\equiv \langle a^{\dagger}_0 a_0\rangle_{\epsilon}-|\alpha|^2\approx \Delta\mathcal{N}_{\epsilon}$ with $\Delta\mathcal{N}_{\epsilon}$ given by Eq.~\eqref{eq:weak_n_sigma}, see the App.~\ref{App:WignerCentered}.  

The expectation value of the number operator $a^{\dagger}_0 a_0$ can be equivalently found using either its Glauber-Sudarshan, Husimi, or Wigner quasi-probability distributions~\cite{Walls2008representations}. The latter is particularly interesting as its negative values certify the quantum nature of the electromagnetic field~\cite{Kenfack2004} and may witness contexuality~\cite{Spekkens2008,Booth2021, Haferkamp2021}, exactly as anomalous weak values~\cite{Aharanov2002,Jordan2008,Pusey2014}. For this reason here we derive the conditional Wigner function of the mode of frequency $\omega_0$:
 \begin{align}\label{eq:WFf}
    &\mathcal{W}_{\epsilon}(\mu)\equiv\frac{1}{\pi^{2}}\int\text{d}^{2}\lambda e^{ -\lambda\mu^{*}+\lambda^{*}\mu} \text{Tr}\left\lbrace e^{-\lambda^{*}a_0+\lambda a_0^{\dagger}}\ket{\psi_{\epsilon}(\tau)}\bra{\psi_{\epsilon}(\tau)}\right\rbrace\\ \nonumber
    &=\frac{2e^{-2|\mu-\alpha|^2}}{\pi P_{\epsilon}(\tau)}\left[\left(P_{\epsilon}(\tau)-|\tilde{f}_{\epsilon}^{(1)}(\tau,0)|^2\right)-|\tilde{f}_{\epsilon}^{(1)}(\tau,0)|^2(1-4 |\mu-\alpha|^2)\right.\\ \nonumber
    &\left.-4 \text{Re}\left\lbrace \tilde{f}^{(1)*}_{\epsilon}(\tau,0)f^{(0)}_{\epsilon}(\tau)\left(\mu-\alpha \right)\right\rbrace +..\right],
\end{align}
where $\tilde{f}_{\epsilon}^{(1)}(\tau,\omega_k-\omega_0)=\sqrt{\frac{1}{\varrho}}\int_0^{\tau} dt f^{(1)}_{\epsilon}(\tau,t)e^{-i(\omega_k-\omega_0)t}$, and $|\tilde{f}_{\epsilon}^{(1)}(\tau,0)|^2/P_{\epsilon}(\tau)$ is the probability that the qubit spontaneously emits one photon of frequency $\omega_0$. In the limit of monochromatic emission, no photon with frequency $\omega_k\neq \omega_0$ is emitted, hence $|\tilde{f}_{\epsilon}^{(1)}(\tau,0)|^2=p^{(1)}(\tau)$. The explicit derivation of Eq.~\eqref{eq:WFf}, including the terms coming from the multi-photon emission (ellipsis), is given in the App.~\ref{App:WignerCentered}. 

Figure~\ref{fig:fig2} shows the main result of this section: the conditional Wigner function $\mathcal{W}_{\epsilon}(\mu)$ takes negative values when $\Delta\mathcal{N}_{\epsilon}$ takes anomalous values. Figure~\ref{fig:fig2}(a) shows the values of $\Delta \mathcal{N}_{\epsilon}$ varying the gate's angle $\theta$ in $[0,\pi]$. While $\Delta \mathcal{N}_{e}$ remains between 0 and -1, $\Delta \mathcal{N}_{g}$ takes anomalous values (smaller than -1) for some gate's angle. Figures~\ref{fig:fig2}(d-e) show the Wigner function's negativity, $N(\mathcal{W}_{\epsilon})\equiv\int d^2\mu |\mathcal{W}_{\epsilon}(\mu)|-1$,~\cite{Kenfack2004}. This quantity is zero when $\mathcal{W}_{\epsilon}(\mu)$ is non-negative for every $\mu$, and bigger than zero otherwise. The plot shows $N(\mathcal{W}_e)$ is zero for every value of $\theta$ in the range $[0,\pi]$, see Fig.~\ref{fig:fig2}(d), while $N(\mathcal{W}_g)$ is non-zero for all the values of $\theta$ such that $\Delta \mathcal{N}_{g}$ is smaller than $-1$, see Fig.~\ref{fig:fig2}(e).

\section{Conclusion}\label{Conclusions}
We presented an anatomical study of a single-qubit gate implemented with a 1D atom driven by a coherent field at resonance. The scattered field is continuously monitored via heterodyne and post-selected over the outcomes of a qubit's projective measurement. Using a collision model, we derived the analytical expression of the field's Husimi-Q function conditioned on the outcomes of the qubit's measurement. The conditional Husimi grants access to all the moments of the post-selected heterodyne distribution, i.e. the field's weak values. In particular, we used the conditional Husimi-Q function to derive the weak value of the field's energy change. As recently observed in~\cite{Stevens2022}, this quantity can exceed by far the single quantum, hence reaching anomalous values. Using the analytical expression of the atom-field wavefunction, we derived the field's conditional Wigner functions. Then, we showed that, as expected from general foundational results~\cite{Pusey2014}, anomalous weak values of the energy change correspond to non-zero Wigner function's negativity.

\section*{Acknowledgments}

We warmly thank Mattia Walschaers for his helpful advices. We gratefully acknowledge financial support from the European Union\textquoteright s Horizon 2020 Researchand innovation Programme under the Marie Sklodowska-Curie Grant Agreement No. 861097, the Foundational Questions Institute Fund (Grant No. FQXi-IAF19-01 and Grant No. FQXi-IAF19-05), the John Templeton Foundation (Grant No. 61835), the ANR Research Collaborative Project “Qu-DICE” (Grant No. ANR-PRC-CES47). 

%\emph{Acknowledgements - }
%The authors acknowledge fruitful discussions with ...

\appendix
%\section{Appendix}
\section{Explicit expression of the field's wavefunction}\label{App:FieldsWavefunction}
Explicit expressions of the functions $f^{(j)}_{\epsilon}(\tau,\textbf{t}_j)$, for the joint system's initial state being $\ket{g, \alpha}$, have been derived in \cite{Maffei2022}. Here we report them for the sake of completeness.
\begin{align}\nonumber
&f^{(0)}_{g}(t)=e^{-\gamma t/4}\left[\cos (\Omega' t/2)+\sin(\Omega' t/2)(\gamma)/(2\Omega') \right],\\ \nonumber
&f^{(0)}_{e}(t)=e^{-\gamma t/4}\sin(\Omega' t/2)\Omega/\Omega'
\end{align}
The amplitude of the emitted photons read:
\begin{align}\nonumber
f^{(1)}_{\epsilon}(\tau,t)&=\sqrt{\gamma} f^{(0)}_{\epsilon}(\tau-t) e^{-i\omega_0 t} f^{(0)}_{e}(t),\\ \nonumber
f^{(j>1)}_{\epsilon}(\tau,\textbf{t}_j)&=(\sqrt{\gamma})^j f^{(0)}_{\epsilon}(\tau-t_{j}) e^{-i\omega_0 t_{j}} \times \\ \nonumber &\left[ \prod_{i=2}^{j} f^{(0)}_{e}(t_{i}-t_{i-1})e^{-i\omega_0 t_{i-1}}\right] f^{(0)}_{e}(t_1).
\end{align}
where $\Omega'=\sqrt{(\Omega)^2-\gamma^2/4}$.

%\section{Explicit derivation of the field's Husimi function}
%Here we give the explicit derivation of the Husimi function in Eq.~\eqref{eq:QFEfinal}. The Husimi function in Eq.~\eqref{eq:QF} can be rewritten as:
%\begin{align}\label{eq:QFE}
    %\mathcal{Q}^{(n)}_{\epsilon}(s)=\frac{1}{\pi P_{\epsilon}(\tau)}\text{Tr}\lbrace E_{\epsilon}(\tau,t_{n+1})\Pi^{(n)}_{s}\rho(t_{n+1})\Pi^{(n)}_{s}\rbrace,
%\end{align}
%where $E_{\epsilon}(\tau,t)\equiv U^{\dagger}(\tau-t)\ket{\epsilon}\bra{\epsilon}U(\tau-t)$. Plugging Eq.~\eqref{eq:qevol} in Eq.~\eqref{eq:QFE}, and expanding $U^{(n)}$ at the second order in $\Delta t$, we find:
%\begin{align}\nonumber
    %&\mathcal{Q}^{(n)}_{\epsilon}(s)=\frac{\text{Tr}\left\lbrace E_{\epsilon}(\tau,t_{n+1})\bra{s_n}U^{(n)}\ket{\alpha_n}\rho_{q}(t_{n})\bra{\alpha_n}U^{(n)\dagger}\ket{s_n}\right\rbrace}{\pi P_{\epsilon}(\tau)}\\ \nonumber &=\text{Tr} \left\lbrace E_{\epsilon}(\tau,t_{n+1})\left[\rho_{q}(t_{n+1})+\gamma\Delta t (|\alpha_n-s|^2-1)\sigma\rho_{q}(t_n)\sigma^{\dagger}\right.\\ \nonumber &\left.\left.\left. +\sqrt{\gamma\Delta t}\left((\alpha_n-s)^*\sigma \rho_{q}(t_n)+(s-\alpha_n) \rho_{q}(t_n)\sigma^{\dagger}\right)\right]\right\rbrace \frac{e^{-|s-\alpha_n|^2}}{\pi P_{\epsilon}(\tau)}.
%\end{align}
%Noticing that $\text{Tr}\lbrace E_{\epsilon}(\tau,t_{n+1})\sigma \rho_{q}(t_n)\rbrace/P_{\epsilon}(\tau)$ is the weak value of $\sigma$, written in discrete-time notation, and that $\text{Tr}\lbrace E_{\epsilon}(\tau,t_{n+1}) \rho_{q}(t_{n+1})\rbrace=P_{\epsilon}(\tau)$, we find Eq.~\eqref{eq:QFEfinal}.

\section{Alternative derivation of the Husimi function}\label{App:HusimiDerivation}

The field's wavefunction in Eq.\eqref{eq:Fwavefunction} can be rewritten in terms of discrete-temporal modes:
\begin{align}\label{eq:Dwf}
    \ket{\psi_{\epsilon}(\tau)}=\frac{\bigotimes_{n}\mathcal{D}^{(n)}(\alpha_n)\left(f^{(0)}_{\epsilon}(\tau)-\sum_{n=0}^{N-1}\sqrt{\Delta t} f^{(1)}_{\epsilon}(\tau,t_n)e^{-i\omega_0 t_n }b^{\dagger}_n\right)+..}{\sqrt{P_{\epsilon}(\tau)}}\ket{0}
\end{align}
where $N=\tau/\Delta t$.

The conditional Husimi function can be equivalently computed from the wavefunction~\eqref{eq:Dwf}, including also the components arising from the spontaneous emission of 2-photons. The components with $j>2$ are irrelevant in the present study, as they can be neglected in the typical gate regime $(\gamma\ll \Omega)$, but they can be included following a conceptually analogous derivation. $\mathcal{Q}_{\epsilon}^{(n)}(s)$ can be rewritten as:
\begin{align}\label{eq:QFDD}
    &\mathcal{Q}^{(n)}_{\epsilon}(s)\equiv \frac{1}{\pi}\text{Tr}\lbrace \Pi^{(n)}_{s-\alpha_n}\eta^{(n)}_{\epsilon}\rbrace
\end{align}
where $\eta_{\epsilon}^{(n)}$ is reduced density matrix of the mode $b_n$ written in the displaced reference frame, and $\Pi_{s-\alpha_n}$ projects it over the coherent state of amplitude $s-\alpha_n$. The field state in the displaced reference frame (now including the components with 2-photon emitted) reads:
\begin{align}\nonumber
    \ket{\phi_{\epsilon}}&=\frac{1}{\sqrt{P_{\epsilon}(\tau)}}\left(f^{(0)}_{\epsilon}(\tau)-\sum_{n}\sqrt{\Delta t} f^{(1)}_{\epsilon}(\tau,t_n)e^{-i\omega_0 t_n }b^{\dagger}_n\right. \\\nonumber &\left.+\sum_{n}\sum_{m>n}\Delta t f^{(2)}_{\epsilon}(\tau,t_n,t_m)e^{-i\omega_0 (t_n+t_m) }b^{\dagger}_n b^{\dagger}_{m}\right)\ket{0}. 
\end{align}
Taking the trace over all the discrete-time modes $m\neq n$, we find $\eta_{\epsilon}^{(n)}$:
\begin{align}\label{eq:eta}
    \eta^{(n)}_{\epsilon}&=\text{Tr}_{\otimes m\neq n}\ket{\phi_{\epsilon}}\bra{\phi_{\epsilon}}\\ \nonumber
    &=\frac{1}{P_{\epsilon}(\tau)}\left[\ket{\phi_{\epsilon}^{01}}\bra{\phi_{\epsilon}^{01}}+\sum_{m\neq n} \ket{\phi_{\epsilon}^{12}(m)}\bra{\phi_{\epsilon}^{12}(m)}\right.\\ \nonumber
    &\left.+\Delta t^2 \sum_{m\neq n}\sum_{l\neq n}|f^{(2)}_{\epsilon}(\tau,t_m,t_l)|^2 \ket{0_n}\bra{0_n} \right]
\end{align}
where
\begin{align}\nonumber
    \ket{\phi_{\epsilon}^{01}}=f_{\epsilon}^{(0)}(\tau)\ket{0_n}-\sqrt{\Delta t}f_{\epsilon}^{(1)}(\tau,t_n)e^{-i\omega_0 t_n}\ket{1_n},
\end{align}
and
\begin{align}\nonumber
    &\ket{\phi_{\epsilon}^{12}(m)}=-\sqrt{\Delta t} f_{\epsilon}^{(1)}(\tau,t_m)e^{-i \omega_0 t_n}\ket{0_n}\\ \nonumber
    &+\Delta t (f_{\epsilon}^{(2)}(\tau,t_n,t_{m})+f_{\epsilon}^{(2)}(\tau,t_m,t_{n}))e^{-i\omega_0 (t_n +t_m)}\ket{1_n}.
\end{align}
Plugging Eq.~\eqref{eq:eta} in Eq.~\eqref{eq:QFDD}, we obtain:
\begin{align}\label{eq:QFsup}
    &\mathcal{Q}^{(n)}_{\epsilon}(s) =\frac{e^{-|s-\alpha_n|^2}}{\pi P_{\epsilon}(\tau)} \left[P_{\epsilon}(\tau)\right.\\ \nonumber
    &+\Delta t\left( |f^{(1)}_{\epsilon}(\tau,t_n)|^2+\sum_{m>n}\Delta t |f^{(2)}_{\epsilon}(\tau,t_n,t_m)|^2\right.
    \\ \nonumber 
    &\left.+\sum_{m<n}\Delta t |f^{(2)}_{\epsilon}(\tau,t_m,t_n)|^2\right) (|\alpha_n-s|^2-1)\\ \nonumber
    &-2\sqrt{\Delta t}\text{Re}\left\lbrace e^{i\omega_0 t_n} \left(s-\alpha_n \right) \left(f^{(1)}_{\epsilon}(\tau,t_n)f^{(0)}_{\epsilon}(\tau)\right.\right.
    \\ \nonumber
    &
    +\sum_{m>n}\Delta t f^{(2)}_{\epsilon}(\tau,t_n,t_m) f^{(1)}_{\epsilon}(\tau,t_m)\\ \nonumber
    &\left.\left.+\sum_{m<n}\Delta t f^{(2)}_{\epsilon}(\tau,t_m,t_n) f^{(1)}_{\epsilon}(\tau,t_m) \right)\right\rbrace]
\end{align}

\begin{figure}[!htb]
\includegraphics[width = 0.5\textwidth]{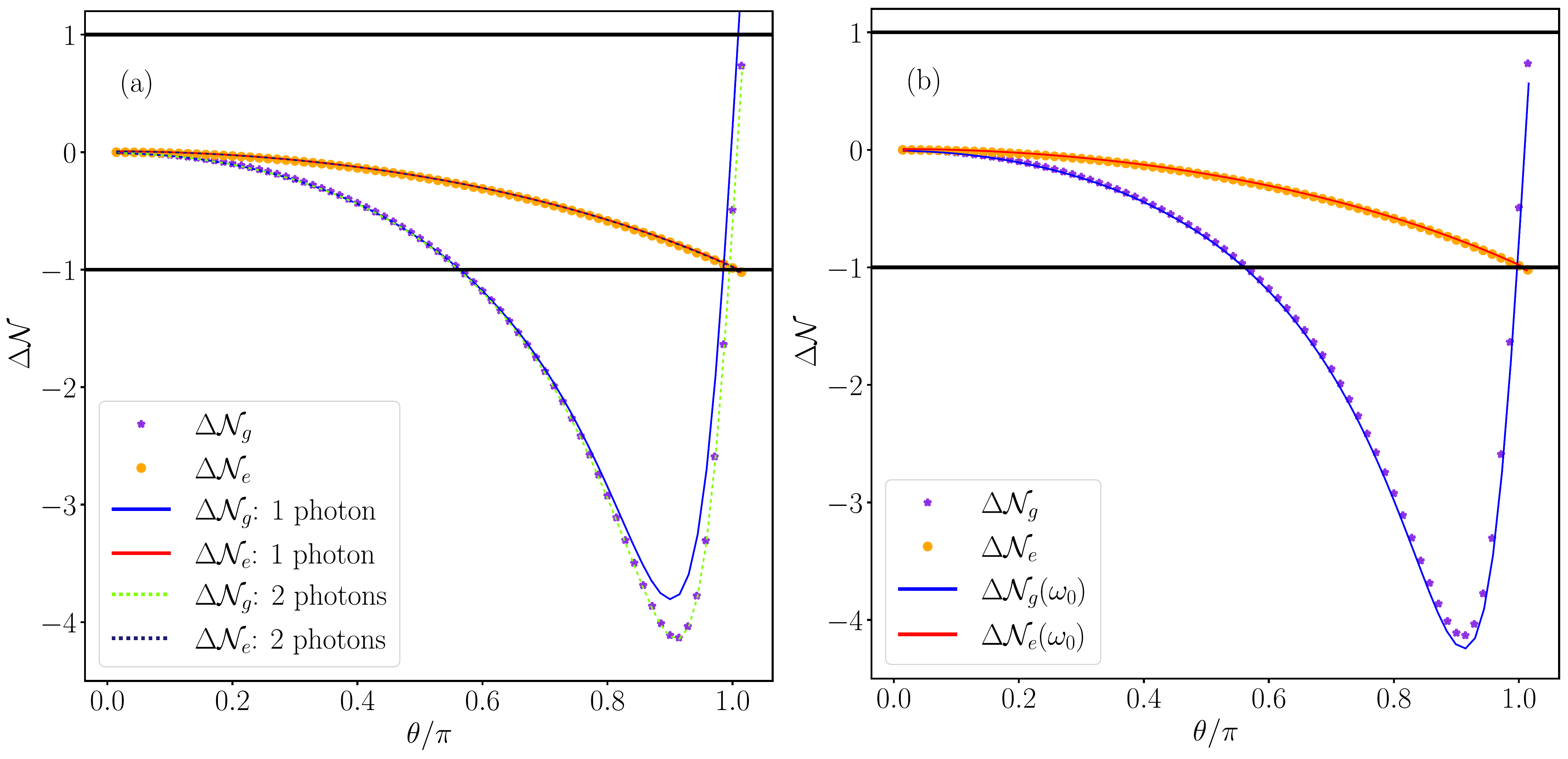}
\caption{(a) Comparison between $\Delta \mathcal{N}_{\epsilon}$ computed using different truncation of the field's wavefunction. The gate considered has $\gamma \tau =3/40$, and the gate's angle $\theta$ varies on the $x-$axis in units of $\pi$. The exact values of $\Delta \mathcal{N}_{\epsilon}$ (solid lines) have been computed from Eq.~\eqref{eq:weak_n_sigma}, by numerical integration of the qubit's forward and backwards Lindblad master equation. The values computed truncating the field's wavefunction at the component with 2 photons emitted (dashed lines) match the exact solutions, while those computed with the truncation at 1 photon emitted (dots) diverge from the exact result after $\theta\approx 0.8\pi$. (b) Comparison between the change in the total field's number of excitations, $\Delta \mathcal{N}_{\epsilon}$, and the change of the field's number of excitations with frequency $\omega_0$, $\Delta \mathcal{N}_{\epsilon}(\omega_0)$, in the considered regime the mismatch among the two data sets is negligible, i.e. the output field can be considered as monochromatic.}
\label{fig:fig3}
\end{figure}
Using this expression we find the following average change in the photon number:
\begin{align}\label{eq:weakn2}
  &\Delta \mathcal{N}_{\epsilon}=\frac{1}{P_{\epsilon}(\tau)}\left(p^{(1)}_{\epsilon}(\tau)+p^{(2)}_{\epsilon}(\tau)\right.\\ \nonumber
  &-2\text{Re}\left\lbrace f^{(0)}_{\epsilon}(\tau)\int_{0}^{\tau} dt \alpha^{*}_t e^{-i\omega_0 t}f^{(1)}_{\epsilon}(\tau,t)\right\rbrace\\ \nonumber
  &-2\text{Re}\left\lbrace \int_{0}^{\tau} dt \int_{t}^{\tau} dt' f^{(1)}_{\epsilon}(\tau,t') \alpha^{*}_t e^{-i\omega_0 t}f^{(2)}_{\epsilon}(\tau,t,t')\right. \\ \nonumber
  &\left.\left.+\int_{0}^{\tau} dt \int_{0}^{t} dt' f^{(1)}_{\epsilon}(\tau,t') \alpha^{*}_t e^{-i\omega_0 t}f^{(2)}_{\epsilon}(\tau,t',t)\right\rbrace \right).
\end{align}
Figure~\ref{fig:fig3}(a) shows that, in the regime considered, i.e. $\gamma^{-1}=8\tau$, Eq.~\eqref{eq:weakn2} matches the exact expression of $\Delta\mathcal{N}_{\epsilon}$ given in Eq.~\eqref{eq:weak_n_sigma}.

\section{Wigner function of the mode with frequency $\omega_0$}\label{App:WignerCentered}
Here we give the explicit derivation of the Wigner function in Eq.~\eqref{eq:WFf}, including also the components arising from the spontaneous emission of 2-photons. The components with $j>2$ are irrelevant in the present study, as they can be neglected in the typical gate regime $(\gamma\ll \Omega)$, but they can be included following a conceptually analogous derivation.
Let us first rewrite the field's wavefunction, Eq.~\eqref{eq:Fwavefunction}, in terms of the frequency modes using the inverse transformation of Eq.~\eqref{eq:a}:
\begin{align}\label{eq:Fwf}
    \ket{\psi_{\epsilon}}=\frac{\mathcal{D}(\alpha)\left(f^{(0)}_{\epsilon}(\tau)-\sum_{k}\tilde{f}^{(1)}_{\epsilon}(\tau,\omega_k-\omega_0) a^{\dagger}_k+..\right)\ket{0}}{\sqrt{P_{\epsilon}(\tau)}},
\end{align}
where $\tilde{f}_{\epsilon}^{(1)}(\tau,\omega_k-\omega_0)=\sqrt{\frac{1}{\varrho}}\int_0^{\tau} dt f^{(1)}_{\epsilon}(\tau,t)e^{-i(\omega_k-\omega_0)t}$, and $a_k$ destroys a photon of frequency $\omega_k$.

Now le us notice that $\mathcal{W}_{\epsilon}(\mu)$ can be rewritten as:
\begin{align}\label{eq:WFfD}
   \mathcal{W}_{\epsilon}(\mu)&\equiv\frac{1}{\pi^{2}}\int\text{d}^{2}\lambda e^{ -\lambda(\mu-\alpha)^{*}+\lambda^{*}(\mu-\alpha)} \text{Tr}\left\lbrace e^{-\lambda^{*}a_0+\lambda a_0^{\dagger}}\zeta_{\epsilon}\right\rbrace
\end{align}
where $\zeta_{\epsilon}$ is reduced density matrix of the mode of frequency $\omega_0$ written in the displaced reference frame. In order to obtain $\zeta_{\epsilon}$ let us first write the field state (Eq.\eqref{eq:Fwf}) in the displaced reference frame, including the components with 2-photon emitted, in terms of frequency modes:
\begin{align}\label{eq:Fwf2}\nonumber
    \ket{\phi_{\epsilon}}=\frac{1}{\sqrt{P_{\epsilon}(\tau)}}\left(f^{(0)}_{\epsilon}(\tau)-\sum_{k}\tilde{f}^{(1)}_{\epsilon}(\tau,\omega_k-\omega_0) a^{\dagger}_k\right.\\ \nonumber
    \left.+\sum_{k}\sum_{k'}\tilde{f}^{(2)}_{\epsilon}(\tau,\omega_k-\omega_0,\omega_{k'}-\omega_0) a^{\dagger}_k a^{\dagger}_{k'}\right)\ket{0},
\end{align}
where
\begin{widetext}
\begin{equation*}
\tilde{f}_{\epsilon}^{(2)}(\tau,\omega_k-\omega_0,\omega_{k'}-\omega_0)=\frac{1}{\varrho}\int_0^{\tau} dt e^{-i(\omega_k-\omega_0)t} \int_{0}^{t} dt' f^{(2)}_{\epsilon}(\tau,t',t)e^{-i(\omega_k-\omega_0)t'}.
\end{equation*}
\end{widetext}
Now we can define the reduced density matrix of the mode of frequency $\omega_0$ taking the trace over the modes with $k\neq0$.
\begin{align}
    \zeta_{\epsilon}&=\text{Tr}_{\otimes k\neq 0}\ket{\phi_{\epsilon}}\bra{\phi_{\epsilon}}\nonumber\\ \nonumber
    &=\frac{1}{P_{\epsilon}(\tau)}\left[\ket{\varphi_{\epsilon}^{012}}\bra{\varphi_{\epsilon}^{012}}+\sum_{k\neq 0} \ket{\varphi_{\epsilon}^{12}(k)}\bra{\varphi_{\epsilon}^{12}(k)}\right.\\ \nonumber
    &\left.+\sum_{k'\neq 0}\sum_{k\neq 0}|\tilde{f}^{(2)}_{\epsilon}(\tau,\omega_k-\omega_0,\omega_{k'}-\omega_0)|^2 \ket{0}\bra{0} \right],
\end{align}
where
\begin{align}
    \ket{\varphi_{\epsilon}^{012}}=f_{\epsilon}^{(0)}(\tau)\ket{0}-\tilde{f}_{\epsilon}^{(1)}(\tau,0)\ket{1}+\sqrt{2}\tilde{f}_{\epsilon}^{(2)}(\tau,0,0)\ket{2},\nonumber
\end{align}
and
\begin{align}
    &\ket{\varphi_{\epsilon}^{12}(k)}=- \tilde{f}_{\epsilon}^{(1)}(\tau,\omega_k-\omega_0)\ket{0}\nonumber\\ \nonumber
    &+(\tilde{f}_{\epsilon}^{(2)}(\tau,0,\omega_k-\omega_0)+\tilde{f}_{\epsilon}^{(2)}(\tau,\omega_k-\omega_0,0))\ket{1}.
\end{align}
 The matrix $\zeta_{\epsilon}$ can be simplified:
\begin{align}\label{eq:zeta}\nonumber
    \zeta_{\epsilon}&=\frac{1}{P_{\epsilon}(\tau)}[ \left( P_{\epsilon}(\tau)-|\tilde{f}_{\epsilon}^{(1)}(\tau,0)|^2-2|\tilde{f}_{\epsilon}^{(2)}(\tau,0,0)|^2\right) \ket{0}\bra{0}\\
    &-\tilde{f}_{\epsilon}^{(1)*}(\tau,0)\tilde{f}_{\epsilon}^{(0)}(\tau)\ket{0}\bra{1}+ h.c.\\ \nonumber
    &+|\tilde{f}_{\epsilon}^{(1)}(\tau,0)|^2\ket{1}\bra{1} +\sqrt{2}\tilde{f}_{\epsilon}^{(2)*}(\tau,0,0)\tilde{f}_{\epsilon}^{(0)}(\tau)\ket{0}\bra{2}+ h.c.
    \\ \nonumber
    & -\sqrt{2}\tilde{f}_{\epsilon}^{(2)*}(\tau,0,0)\tilde{f}_{\epsilon}^{(1)}(\tau,0)\ket{1}\bra{2}+ h.c.\\ \nonumber
    &+2 |\tilde{f}_{\epsilon}^{(2)}(\tau,0,0)|^2 \ket{2}\bra{2}]
\end{align}
where we neglected the terms containing $\tilde{f}_{\epsilon}^{(2)}(\tau,0,\omega_k-\omega_0)$ or $\tilde{f}_{\epsilon}^{(2)}(\tau,\omega_k-\omega_0,0)$ with $k\neq0$ since they correspond to the unlikely emission of two photons with frequency different from $\omega_0$. The Wigner function can be now computed analytically plugging Eq.\eqref{eq:zeta} in Eq.\eqref{eq:WFfD}:
\begin{align}\label{eq:WFf2}\nonumber
    &\mathcal{W}(\mu)=\frac{2e^{-2|\mu-\alpha|^2}}{\pi P_{\epsilon}(\tau)}\left[\left( P_{\epsilon}(\tau)-|\tilde{f}_{\epsilon}^{(1)}(\tau,0)|^2-2|\tilde{f}_{\epsilon}^{(2)}(\tau,0,0)|^2\right)\right.\\ \nonumber
    &-|\tilde{f}_{\epsilon}^{(1)}(\tau,0)|^2 L_{1}(4\vert\mu-\alpha\vert^{2})+2|\tilde{f}_{\epsilon}^{(2)}(\tau,0,0)|^2 L_{2}(4\vert\mu-\alpha\vert^{2})\\ \nonumber
    &-8 \text{Re}\lbrace\tilde{f}_{\epsilon}^{(2)*}(\tau,0,0)\tilde{f}_{\epsilon}^{(1)}(\tau,0)(\mu-\alpha)^*\rbrace(2|\mu-\alpha|^2-1)\\ \nonumber
    &+8 \text{Re}\lbrace\tilde{f}_{\epsilon}^{(2)*}(\tau,0,0)f_{\epsilon}^{(0)}(\tau)(\mu-\alpha)^2\rbrace\\
    &\left.-4 \text{Re}\lbrace f^{(0)}_{\epsilon}(\tau)\tilde{f}^{(1)*}_{\epsilon}(\tau,0)\left(\mu-\alpha \right)\rbrace\right],
\end{align}
where $L_{n}(x)$ are the Laguerre polynomials. The change in the number of excitations of the field with frequency $\omega_0$, for the two post-selections, can be computed from the corresponding Wigner functions:
\begin{align}\label{eq:NFf}
    &\Delta \mathcal{N}_{\epsilon}(\omega_0)=\int d^{2}\mu \left(|\mu|^2-\frac{1}{2}\right) \mathcal{W}_{\epsilon}(\mu)-|\alpha|^2\\ \nonumber&=\frac{1}{P_{\epsilon}(\tau)}\left(|\tilde{f}^{(1)}_{\epsilon}(\tau,0)|^2-2\text{Re}\lbrace \alpha \tilde{f}^{(1)*}_{\epsilon}(\tau,0) f^{(0)}_{\epsilon}(\tau)\rbrace+..\right).
\end{align}
When the scattered field can be considered as monochromatic $\Delta \mathcal{N}_{\epsilon}(\omega_0)\approx \Delta \mathcal{N}_{\epsilon}$, see Fig.~\ref{fig:fig3}(b).

\end{document}